\documentclass[twoside]{article}
\usepackage{qic}

\usepackage{amsmath}
\usepackage{graphicx}
\usepackage[caption=false]{subfig}

\usepackage[sort&compress,numbers]{natbib}

\newcommand{\tr}[0]{{\mathrm{tr}}}

\begin{document}

\setlength{\textheight}{8.0truein}    

\runninghead{Many-Body Quantum State Control in the Presence of Environmental Noise}
            {Zara Yu and Da-Wei Luo}

\normalsize\textlineskip
\thispagestyle{empty}
\setcounter{page}{1}

\copyrightheading{0}{0}{2003}{000--000}

\vspace*{0.88truein}

\alphfootnote

\fpage{1}

\centerline{\bf
Many-Body Quantum State Control in the Presence of Environmental Noise}

\vspace*{0.37truein}
\centerline{\footnotesize
Zara Yu}
\vspace*{0.015truein}
\centerline{\footnotesize\it Massachusetts Institute of Technology}
\baselineskip=10pt
\centerline{\footnotesize\it{Cambridge, MA 02139, USA}}
\vspace*{0.015truein}
\centerline{\footnotesize\it Department of Physics and Center for Quantum Science and Engineering, Stevens Institute of Technology}
\baselineskip=10pt
\centerline{\footnotesize\it Hoboken, NJ 07030, USA}
\vspace*{10pt}
\centerline{\footnotesize 
Da-Wei Luo}
\vspace*{0.015truein}
\centerline{\footnotesize\it Department of Physics and Center for Quantum Science and Engineering, Stevens Institute of Technology}
\baselineskip=10pt
\centerline{\footnotesize\it Hoboken, NJ 07030, USA}
\vspace*{0.225truein}
\publisher{(received date)}{(revised date)}

\vspace*{0.21truein}

\abstracts{
We consider the quantum state control of a multi-state system which evolves an initial state into a target state. We explicitly demonstrate the control method in an interesting case involving the transfer and rotation of a Schr\"{o}dinger cat state through a coupled harmonic oscillator chain at a predetermined time $T$. We use the gradient-based Krotov's method to design the time-dependent parameters of the coupled chain to find an optimal control shape that will evolve the system into a target state. We show that the prescribed quantum state control can be achieved with high fidelity, and the robustness of the control against generic environment noises is explored. Our findings will be of interest for the optimal control of a many-body open quantum system in the presence of environmental noise.
}{}{}

\vspace*{10pt}

\vspace*{3pt}
\communicate{to be filled by the Editorial}

\vspace*{1pt}\textlineskip    

\section{Introduction}

Quantum state control is crucial for quantum information processing and transmitting quantum information through a quantum network. In this paper, we study how to control the time evolution of a many-body quantum system by designing time-dependent physical parameters that describe the system's internal geometry or configuration.  An interesting classical analogy is that a falling cat can re-orient itself so that it lands on its feet and maximally reduces damage to its body~\cite{Sechzer1984g,McDonald1955n,Gbur2019u,Chryssomalakos2015l}. The controllability of this classical phenomenon correlates with the fact that a cat is not a rigid body~\cite{Montgomery1993w}, but can change the shape of its body and the relative orientations of its body parts, allowing it to rotate without violation of the laws of angular momentum conservation. In the quantum domain, the issue of autonomous control can become more complicated as a quantum deformable body is not a well-studied platform that can easily exhibit quantum control features. To illustrate our methodology, we consider a quantum system which consists of a chain of coupled harmonic oscillators, which we will use to show the transfer and re-orientation of quantum cat states through varying couplings and frequencies for a given control runtime.

The goal of the optimal control for the system considered in this paper is to implement a combination of simultaneous quantum state transfer and cat-state rotation at a predetermined time $T$.  The harmonic oscillator model has been studied extensively~\cite{Buchmann2018k,Brun1999f,Halliwell2003q,Audenaert2002d,Huetal2008,Zhao2014t,Hu1992a}. One of the advantages of the coupled harmonic oscillator model is that it is simple, yet able to exhibit many complex and interesting features of quantum control processes. We will use a gradient-based search method known as Krotov's method to program the control strategies~\cite{Reich2012v,Sklarz2002r,Konnov1999b,Tannor1992h}. Our purpose is to find the optimal control geometry to realize the required state control. Krotov's method is a very versatile and effective control algorithm~\cite{Goerz2019w,Reich2012v,Sklarz2002r,Konnov1999b,Tannor1992h}, applied to many quantum control problems, including quantum gate preparation~\cite{Huang2014e} and quantum state transfer in a spin chain~\cite{Zhang2016f}. One major issue in quantum control problems is the robustness of the control strategy against detrimental noises. To address this issue, we study the robustness of the quantum control strategy in the presence of a colored noisy environment~\cite{Breuer2002a} by using a stochastic Schr\"{o}dinger equation~\cite{Diosi1998,Yu1999a}.

This paper is organized as follows: in Section~\ref{sec_modctrl}, we first introduce the quantum model under consideration as well as the control strategy. We then present the resultant control dynamics in a generic open system setting and discuss the results on some specific cases in Section~\ref{sec_resopen}. We give a conclusion in Section~\ref{sec_sum}.

\section{The model and optimal control}\label{sec_modctrl}

\begin{figure}
  \centering
  \centerline{\epsfig{file=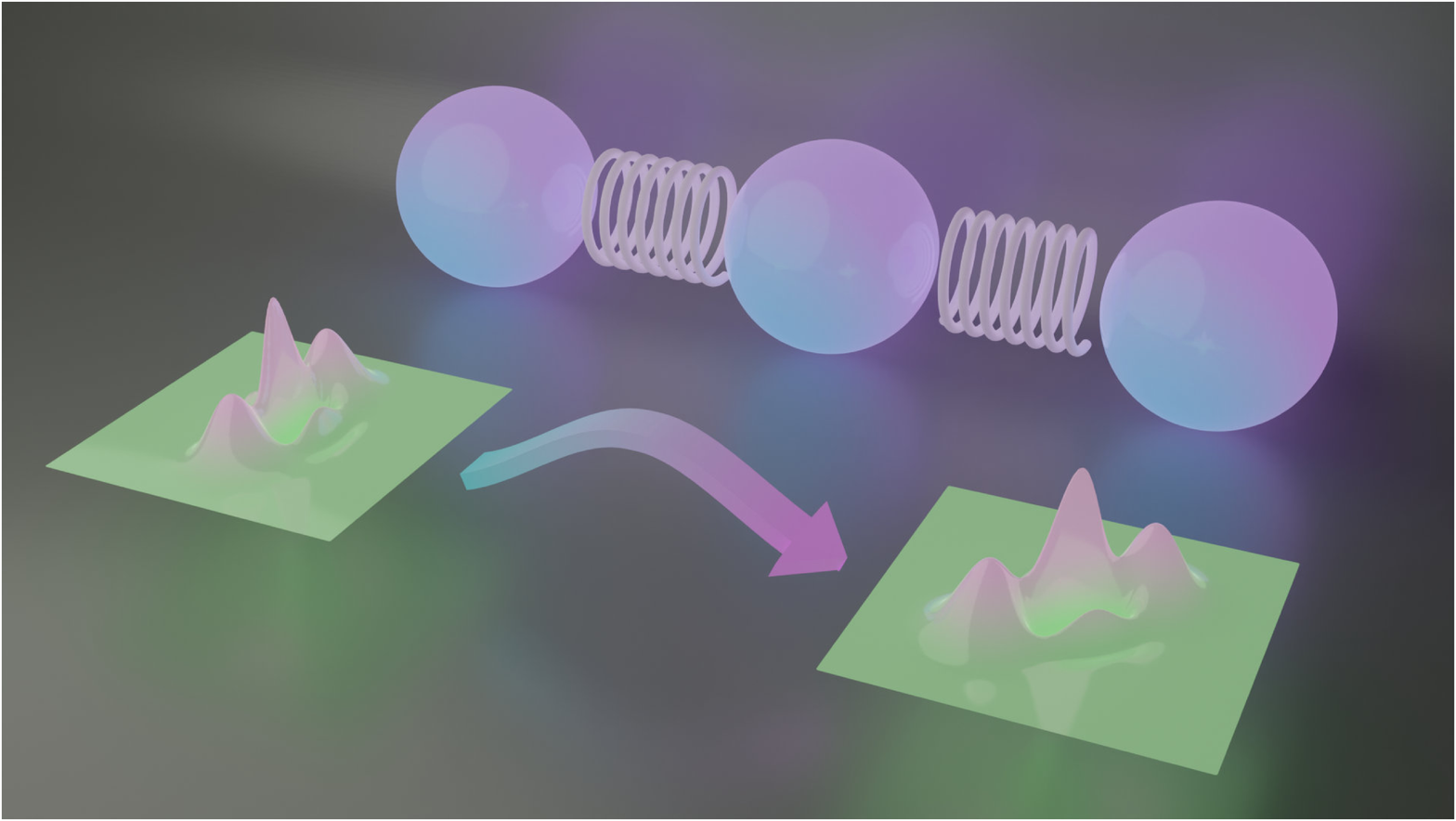, width=8.2cm}} 
  \vspace*{13pt}
  \fcaption{Schematic of the model with a chain of three harmonic oscillators. The control is implemented to transfer a cat state from the first oscillator to the last one in the chain at a given time $T$, where the transferred state may be rotated by a pre-determined angle. For simplicity, we initially assume that the rest of the oscillators are in the vacuum states.}\label{fig_model0}
\end{figure}

We consider a quantum system consisting of $N$ coupled harmonic oscillators in an open-ended chain, coupled by position-position operators at neighboring sites. The model may be represented by a set of bosonic annihilation and creation operators $a_j(a_j^\dagger) = \sqrt{\frac{\omega_j}{2}}(q_j\pm \frac{i}{\omega_j}p_j)$, where $j=1\ldots N$. Dropping the trivial overall constant terms, which only induce overall phase factors, as well as the counter-rotating terms $a_j a_{j+1}$ and $a_j^\dagger a_{j+1}^\dagger$, the Hamiltonian may be written as (setting $\hbar=1$)
\begin{align}
    H(t) &= \sum_{j=1}^N \tilde\omega_j(t) a_j ^\dagger a_j + \sum_{j=1}^{N-1}\tilde k_j(t) \left(a_j ^\dagger a_{j+1} + h.c.\right), \nonumber\\
    &\equiv \sum_{j=1}^N \tilde\omega_j(t) h_{o,j} + \sum_{j=1}^{N-1}\tilde k_j(t) h_{k,j}  \label{eq_hosc},
\end{align}
where the explicitly time-dependent $\tilde\omega_j(t)$ and $\tilde k_j(t)$ would serve as our quantum control parameters, and $h_{o(k),j}$ denotes the oscillator (coupling) Hamiltonian for site $j$. It should be noted that the model may be implemented in many interesting physical systems, such as an array of coupled optical cavities~\cite{Hartmann2008u}. The coupled time-dependent harmonic oscillator model~\cite{Macedo2012t} also appears in various interesting physical systems, such as a cavity with a moving mirror~\cite{Law1994y}, quantum circuit systems~\cite{Zhang2002c}, and charged particles in time-varying magnetic fields~\cite{Menouar2010f}.

We will consider the control of a class of quantum states generally referred to as Schr\"{o}dinger cat states, representing a set of macroscopically distinct superposition states. To be more specific, a cat state here consists of a superposition of two Gaussian wave packets in the opposite direction, with interference patterns in between. It has been found that Schr\"{o}dinger cat states have various applications in quantum technologies, such as coherent state-based quantum information processing and fundamental physics (eg, see \cite{Ralph2003y,Cochrane1999u,Anastopoulos2015g}). Recently, long-lived cat states have been shown to be realizable in atomic ensembles~\cite{Qin2021q}, and a dissipative cat state generation via nonequilibrium pump fields has also been reported~\cite{zynori2021, zynori2022}. An alternative approach has also been reported using shortcuts to adiabaticity~\cite{yh_s2ad}. It should be noted that Schr\"{o}dinger cat states may be generated in quantum optical systems~\cite{Gerry1997d} or Bose-Einstein condensates~\cite{Cirac1998x}.

In order to realize the optimal control of a quantum state at a predetermined time $T$, we will consider the transfer and orientation of the quantum cat states, which is given by $|\psi_c(\alpha) \rangle = \mathcal{N}_\alpha \left(|\alpha \rangle + |-\alpha \rangle\right)$, where $|\alpha\rangle$ is the coherent state represented by the complex number $\alpha$ and $\mathcal{N}_\alpha$ is the normalization factor. The rotation of the cat state is represented by the phase angle $\theta$ of the coherent state parameter, $\alpha = |\alpha|e^{i \theta}$. The control goal here is to transfer the cat state through the chain of oscillators to a target state. As an example, the target state can be a transferred cat state with a predetermined rotation angle $\theta_T$. More precisely, the initial and target state of the quantum cat system are given by
\begin{align}
    |\psi(0) \rangle &= |\psi_c(\alpha) \rangle_1 \otimes |0 \rangle_2 \otimes \ldots \otimes| 0 \rangle_N \\
    |\psi(T) \rangle &=  |0 \rangle_1 \otimes \ldots \otimes| 0 \rangle_{N-1} \otimes |\psi_c(\alpha e^{i \theta_T}) \rangle_{N},
\end{align}
respectively. The crucial question is how to program the quantum system to achieve the required final state within the given time $T$. The control functions ($k_j(t), \omega_j(t)$) that can be used to achieve this goal may be found by a quantum optimal control method. To find the optimal control field, a gradient-based method known as Krotov's method~\cite{Reich2012v,Sklarz2002r,Konnov1999b,Tannor1992h} is used here. Krotov's method has the advantage of being able to monotonically approach the control goal with each iteration, and there is no need for a line search~\cite{Goerz2019w,Hwang2012a}. Krotov's control has been applied to open systems~\cite{Hwang2012a}, the generation of quantum gates~\cite{Chou2015r,Huang2014e} and quantum state transfers~\cite{Zhang2016f}.

Krotov's method starts with a trial solution of the control functions and iteratively optimizes the control functions' shape to minimize a functional $J$ to be explicitly defined below. The resulting control function of the current iteration is then used as the trial function for the next iteration. At the $i$-th iteration, the total Hamiltonian is of the form
\begin{align}
    H^{(i)}(t) &= H_0 + \sum_l \epsilon^{(i)}_l(t) H_l, \label{eq_ctrl}
\end{align}
where $H_0$ is the time-independent uncontrolled part of the Hamiltonian, $H_l$ is the time-independent Hamiltonian describing the controlling strategy, and $\epsilon^{(i)}_l(t)$ is the corresponding time-dependent control function. Krotov's method would minimize the functional $J$,
\begin{align}
    J\left[s, \{\epsilon^{(i)}_l(t)\}\right] = J_T(s) + \sum_l \int_0^T g(\{\epsilon^{(i)}_l(t)\}) \label{eq_j},
\end{align}
where $s=\{| \varphi^{(i)}(t)\}$ is the set of wave functions at the $i$-th iteration, and $\{\epsilon^{(i)}_l(t)\}$ is the set of control functions. The function $J_T$ is the main part of the functional~\eqref{eq_j} and for our purpose, $J_T$ is taken to be the infidelity of the evolved state and the target state,
\begin{align}
    J_T(s) = 1 - |\langle \phi_{f} | \varphi^{(i)}(T) \rangle|^2,
\end{align}
where $|\phi_{f} \rangle$ is the target final state. The function $g$ tracks the running cost of the control fields, and is usually taken in the form of
\begin{align}
    g(\{\epsilon^{(i)}_l(t)\}) = \frac{\lambda_{a,l}}{S_l(t)}(\Delta \epsilon_l^{(i)}(t))^2,
\end{align}
where $\lambda_{a,l}>0$ is an inverse step-size, $\Delta \epsilon_l^{(i)}(t)= \epsilon_l^{(i)}(t) - \epsilon_l^{(i-1)}(t)$ is the difference of the control function between the current and last iteration, and $S_l(t) \in [0,1]$ is an update shape function. 


In the most general case, the functional $J$ above can include an additional functional $h$ that depends on the state's dynamics in each iteration. However, for our case,  $h(s)$ may be omitted here. With this search method, one starts with a trial solution and in the $i$-th iteration, the $l$-th control field is updated according to
\begin{align}
    \Delta \epsilon_l^{(i)}(t) = \frac{S_l(t)}{\lambda_{a,l}} \mathrm{Im} \left[\left\langle \chi^{(i-1)}(t)\left|\frac{\partial H^{(i)}(t)}{\partial \epsilon^{(i)}_l(t)} \right| \varphi^{(i)}(t)\right\rangle\right],
\end{align}
where $H^{(i)}(t)$ is the total Hamiltonian of the $i$-th iteration and $|\chi^{(i-1)}(t) \rangle$ is back-propagated using the Hamiltonian under the previous iteration's control fields with an appropriate boundary condition $|\chi^{(i-1)}(T) \rangle \propto |\phi_f \rangle$, i.e. the target state.

Krotov's method is shown to be monotonically convergent in the continuous time limit. In practice, when the control is discretized, a proper step size and other parameters need to be chosen. This algorithm~\cite{Goerz2019w,Johansson2013r,Johansson2012f} can be a robust method to find the control shapes to drive a quantum system towards a target state.

As an illustrative example, here we consider a chain of $N=3$ harmonic oscillators (see Fig.~\ref{fig_model0}) and the control functions have a constant part as well as a time-dependent part, $\tilde \omega_j(t) = \omega_{j,0} + \omega_{j}(t)$ and $\tilde k_j(t) = k_{j,0} + k_{j}(t)$. Therefore, rewriting the system Hamiltonian into the form of Eq.~\eqref{eq_ctrl}, we have  $5$ control parameters in total,
\begin{align}
    H_0 &= \sum_{j=1}^N \omega_{j,0} h_{o,j} + \sum_{j=1}^{N-1}k_{j,0} h_{k,j}; \nonumber \\
    H_l &= h_{o,l}, \;\epsilon_l(t) = \omega_l(t) \text{ for } l=1,2,3; \nonumber \\
    H_4 &= h_{k, 1}, \;\epsilon_4(t) = k_1(t) \text{, and } H_5 = h_{k, 2}, \;\epsilon_5(t) = k_2(t).
\end{align}

\section{Results and discussions}\label{sec_resopen}

\begin{figure}
    \centering
    \centerline{\epsfig{file=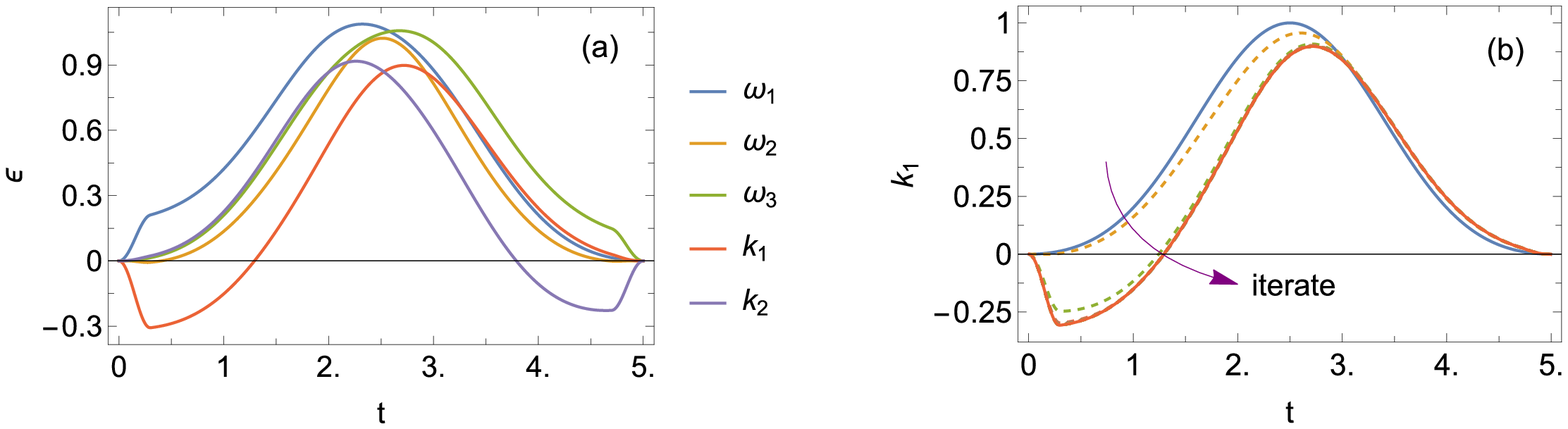, width=0.8\textwidth}} 
    \vspace*{13pt}
    \fcaption{Optimal control shapes for transferring the quantum cat state while rotating it by $90$ degrees as a function of time. Panel (a). The final control shapes for all control fields. Panel (b). The iteration history of the coupling strength $k_1(t)$ as an example, where we started with an initial guess (blue solid lines) as the $1$st iteration, and reached the final shape (red solid line) at the $19th$ iteration with a goal of $10^{-7}$ for the functional $J$. The dashed lines are from iteration $2$ through $18$.}
\label{fig_ctrl_90d}
\end{figure}

We first start with an interesting case where the transferred state will be rotated by $90$ degrees.  For $H_0$, we take $\omega_{j,0}=1$ and $k_{j,0}=0.3$, and we set the target time to be $T=5$. The cat state's parameter is chosen to be $\alpha = 1$. The control shapes for the system is shown in Fig.~\ref{fig_ctrl_90d} (a).  It can be seen from Fig.~\ref{fig_ctrl_90d} (b) that the control function's shape quickly converges to the final shape after only $2$ iterations. For the case where we minimize the functional~\eqref{eq_j} to be $10^{-7}$, the final shape in Fig.~\ref{fig_ctrl_90d} is reached after $19$ iterations. The above analysis shows that a control strategy may be used to realize quantum state transfer and control.  A more generic transfer case with an arbitrary rotation can be treated in a similar fashion, but the iteration times may vary for different target states. 

\begin{figure}
    \centering
    \centerline{\epsfig{file=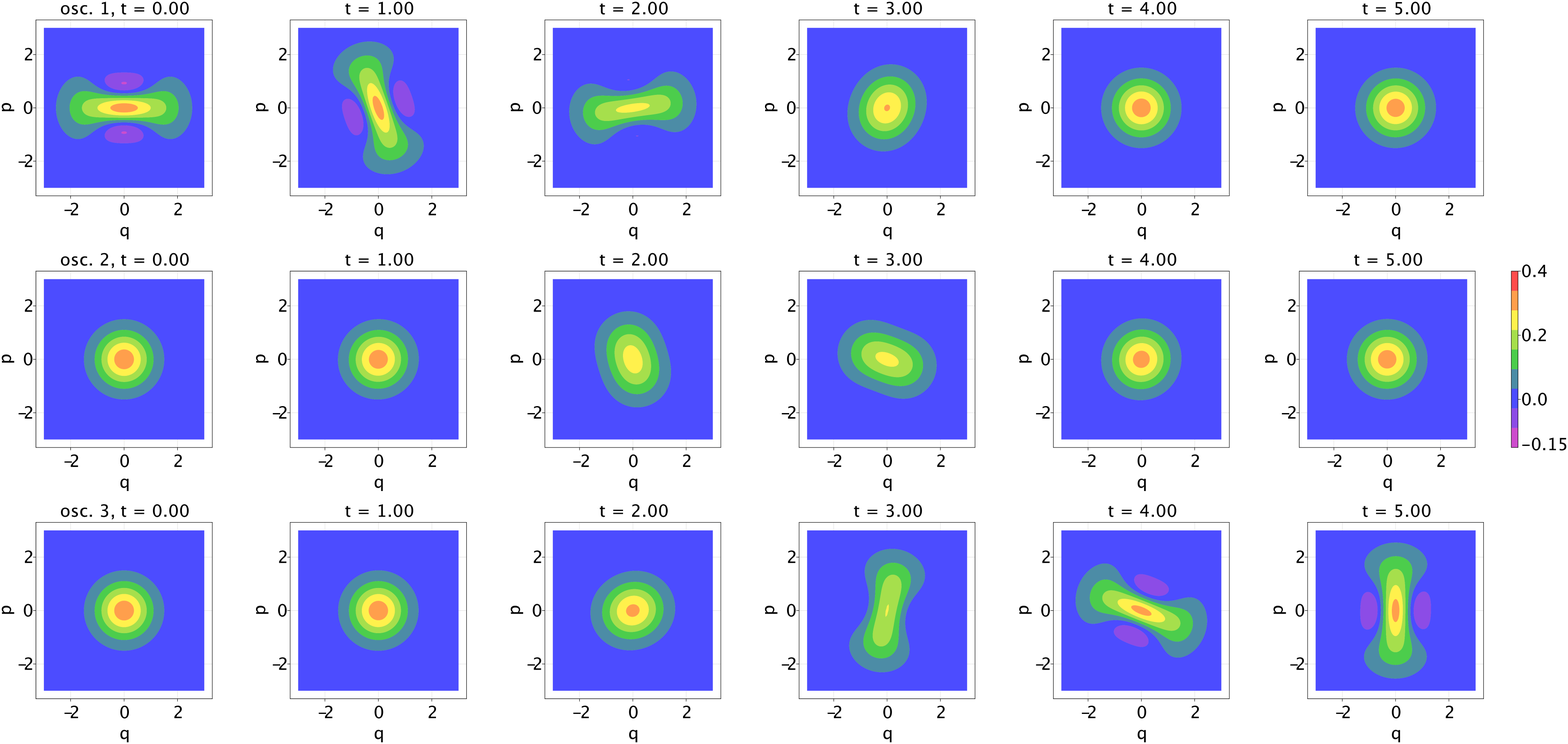, width=0.9\textwidth}}
  \vspace*{13pt}
    \fcaption{The state evolution of the quantum system is plotted. Snapshots of the Wigner function for the quantum states at various times for the three oscillators (osc. 1, 2, 3 on each row), where the system is embedded in a noisy environment.}\label{fig_90d_w}
\end{figure}

In reality, all quantum systems are inevitably coupled to their environment, making them susceptible to environmental noise and decoherence. Therefore, it is interesting to see if a useful control strategy is still available in an open system context. To this end, we take the optimal control field derived above using an ideal closed system evolving unitarily, and study the robustness of the controlled dynamics under the influence of a generic non-Markovian environment. To consistently add noise to the system of interest, we consider a standard model of open system dynamics where a quantum system is embedded in a bosonic environment with total Hamiltonian
\begin{align}
    H_{\rm tot}(t) = H_s(t) + \sum_k \bar\omega_k b_k ^\dagger b_k+ \sum_k \left(g_k L b_k^\dagger + h.c.\right). \label{eq_openhtot}
\end{align}
Here, $H_s(t)$ represents the controlled system Hamiltonian~\eqref{eq_hosc}, $b_k$ is the annihilation operator of the $k$-th bath mode, $\bar\omega_k$ is its frequency, $L$ is the system operator that describes how the system couples to the bath (environment), and $g_k$ is the coupling strength. The system's reduced dynamics can then be obtained by tracing out the environmental degrees of freedom on the combined system which evolves under Eq.~\eqref{eq_openhtot}. While the dynamics of the open system may be described by a Markov approximation under specific conditions such as weak coupling and unstructured bath~\cite{Breuer2002a}, a non-Markovian treatment is generally necessary to account for more intricate memory effects. The non-Markovian dynamics may be solved by a generic quantum-trajectory approach known as the quantum state diffusion (QSD) approach~\cite{Yu1999a,Diosi1998}, where the bath modes are projected onto coherent states, introducing an effective noise such that each trajectory can be determined by a stochastic Schr\"{o}dinger equation. The solutions to the stochastic Schr\"{o}dinger equation are termed quantum trajectories.  The reduced density operator for the open system may then be recovered by an ensemble average of the quantum trajectories. When the system-bath coupling is not in the strong coupling regime, a leading-order master equation may be derived in the form of
\begin{align}
    \frac{d}{d t} \rho_s(t)&=-i \left[H_s(t), \rho_s(t)\right]+\left[L,\rho_s(t)\bar{O}^{(0)\dagger}(t)\right]-\left[L ^\dagger,\bar{O}^{(0)}(t)\rho_s(t)\right]. \label{eq_meq}
\end{align}
The QSD approach introduces a functional derivative operator $\bar{O}$, with respect to the bath state's projection. This operator may be approximated by its leading 
noise-free order $\bar{O}^{(0)}(t)$ in the weak non-Markovian case, which is in turn determined by 
\begin{align}
    &{} \bar{O}^{(0)}(t) = \int_0^t ds \alpha(t,s) O^{(0)}(t,s), \\
    &{} \partial_t O^{(0)}(t,s)=\left[-iH_s(t)  -L ^\dagger \bar{O}^{(0)}(t),O^{(0)}(t,s)\right],
\end{align}
where $\alpha(t,s)=\sum_m |g_m|^2 \exp[-i \bar\omega_m (t-s)]$ is the bath correlation function. Here, we consider a type of noise known as the Ornstein-Uhlenbeck noise~\cite{Diosi1998, qsd_n1, qsd_n2}. This type of noise corresponds to a Lorentzian spectrum of the bath, and gives an exponentially-decaying correlation function $\alpha(t,s) = \gamma \exp \left(-\gamma |t-s|\right)/2$. The $\gamma$ parameter of the correlation function dictates the environmental memory or correlation time~\cite{Yu1999a} $\tau = \gamma^{-1}$. In the limit of $\gamma \rightarrow \infty$, the memory function becomes a $\delta$ function and we get the well-known white noise case such that the dynamics would be memory-less (Markov), whereas for a finite $\gamma$, we have a colored noise.

With this generic open system dynamics treatment, we now take the control functions obtained from the closed system dynamics and proceed to study how robust the control strategy can be when the open system effects are taken into account. Here, we take the system-bath coupling to be $L=\lambda\sum_{i=1}^3q_i$, with the coupling strength $\lambda = 0.1$. In addition, the memory parameter of the correlation function $\alpha$ is chosen to be $\gamma=1.8$. The effect of the environment can be simulated by solving the effective Schr\"{o}dinger equation. This set of parameters is consistent with the weak coupling and weak non-Markovian assumption used here for a noise-free $\bar{O}$ operator.

\begin{figure}
  \centering
  \centerline{\epsfig{file=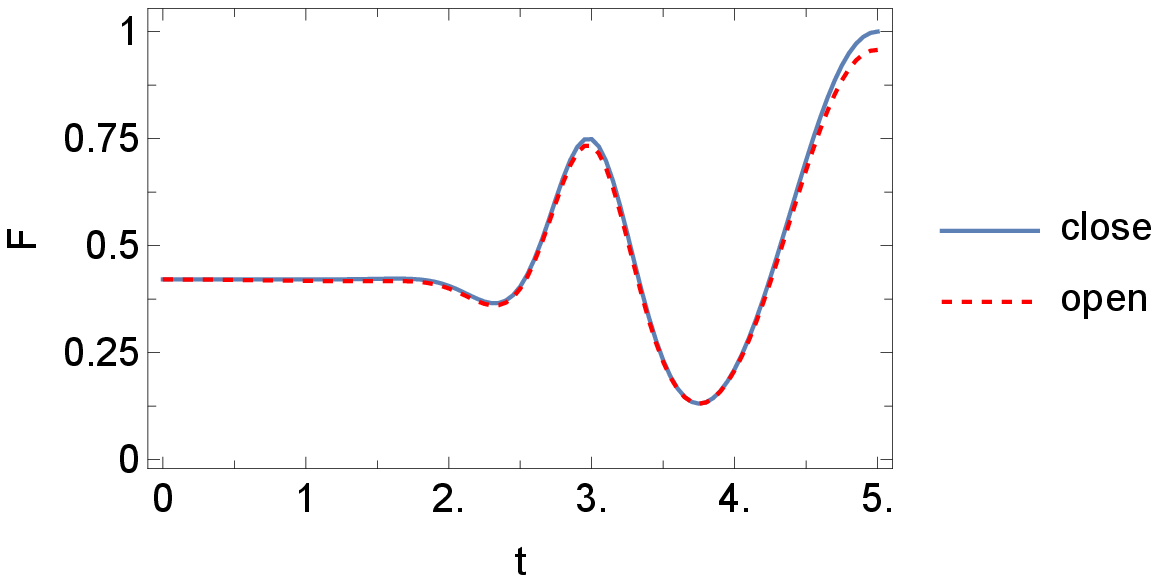, width=8.2cm}} 
  \vspace*{13pt}
  \fcaption{Dynamics of the fidelity between the evolved state and the target state, where the transferred state is engineered to be rotated by $90$ degrees.}\label{fig_f_90d}
\end{figure}

Using the optimal control functions obtained from Krotov's method in the context of a non-Markovian open system, we calculate the dynamics of the Wigner functions for each oscillator in the system. Snapshots of the Wigner function during the evolution are shown in Fig.~\ref{fig_90d_w}, where one can easily visualize how the oscillators' states evolve to the target state in the presence of a dissipative quantum bath. This shows that the control employed here can be quite robust to this type of dissipation. To study the open system effects in detail, we calculate the dynamics of the fidelity of the quantum state against the target state. The fidelity between two (potentially mixed) quantum states $\rho$ and $\sigma$ is given by
\begin{align}
    F = \left[\tr \sqrt{\sqrt{\rho}\sigma\sqrt{\rho}}\right]^2
\end{align}
and takes a real value between $0$ and $1$ such that $F=0$ when the two states are orthogonal and $F=1$ when the two states are identical, up to a global phase. The infidelity is given by $\delta_F=1-F$. We plot the fidelity as a function of time in Fig.~\ref{fig_f_90d}. It can be seen that the open system's fidelity is quite close to the ideal closed system case. As a comparison, the infidelity of the closed system is $\delta_F\approx 1.3 \times 10^{-7}$ at the final time $t_f=5$, while for the open system, we have $\delta_F\approx 0.043$.

\begin{figure}
    \centering
    \subfloat[\label{fig_r45_c}]{%
    \epsfig{file=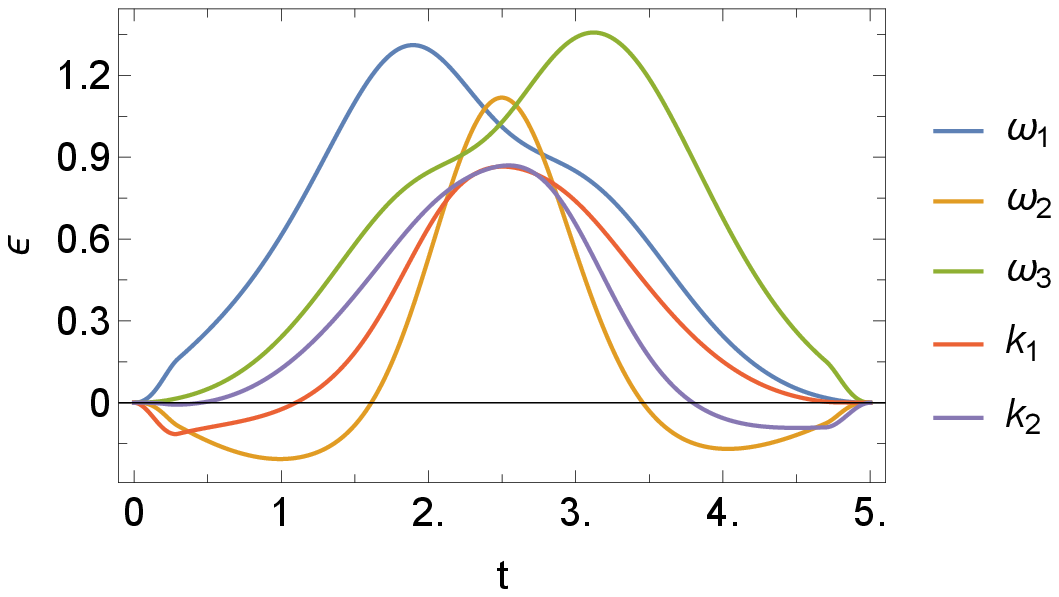, height=3.5cm} 
    }\hspace{9mm}
    \subfloat[\label{fig_r45_f}]{%
    \epsfig{file=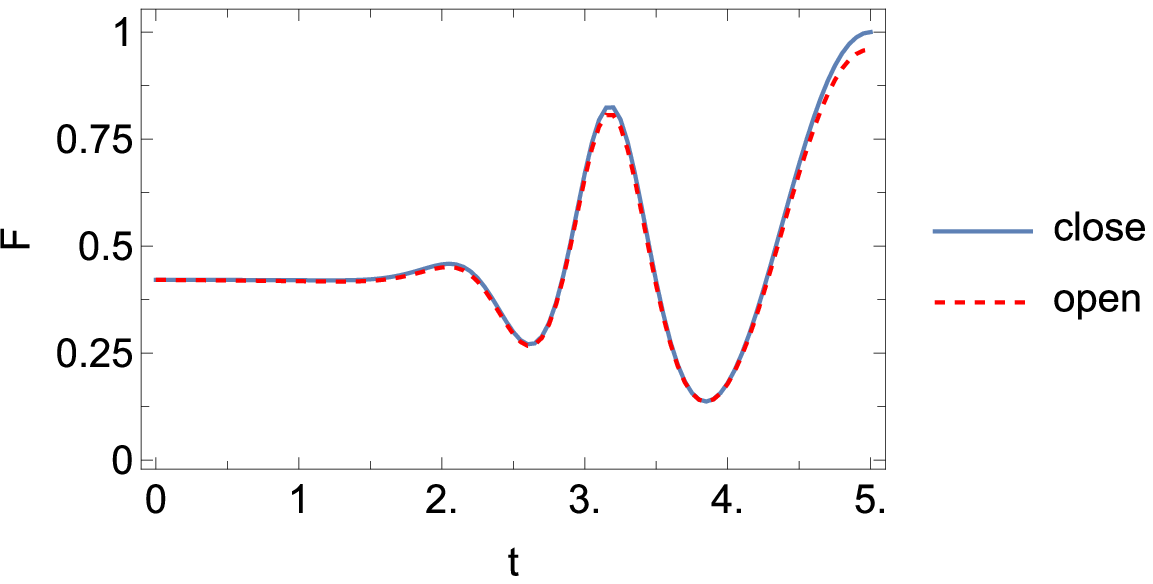, height=3.5cm} 
    }

    \vspace*{13pt}
    \fcaption{Panel (a). Control function shapes for rotating the transferred cat state by $45$ degrees as a function of time. Panel (b). The dynamics of the fidelity between the evolved state and the target state as a function of time for rotating the transferred state by $45$ degrees, under both closed system (blue solid line) and open system dynamics (red dashed line).}\label{fig_r45_cf}
\end{figure}

This control strategy also allows us to choose other angles for the rotation of the transferred cat state. As an example, we show the results for rotating the transferred state at cavity $3$ by $45$ degrees. The final control shapes are shown in Fig.~\ref{fig_r45_c}, and the dynamics of the fidelity between the evolved state and the target state is shown in Fig.~\ref{fig_r45_f}. In this case, we have the infidelity $\delta_F\approx 9.2 \times 10^{-8}$ for the closed system case and $\delta_F\approx 0.042$ for the colored noise case.

\begin{figure}
  \centering
  \centerline{\epsfig{file=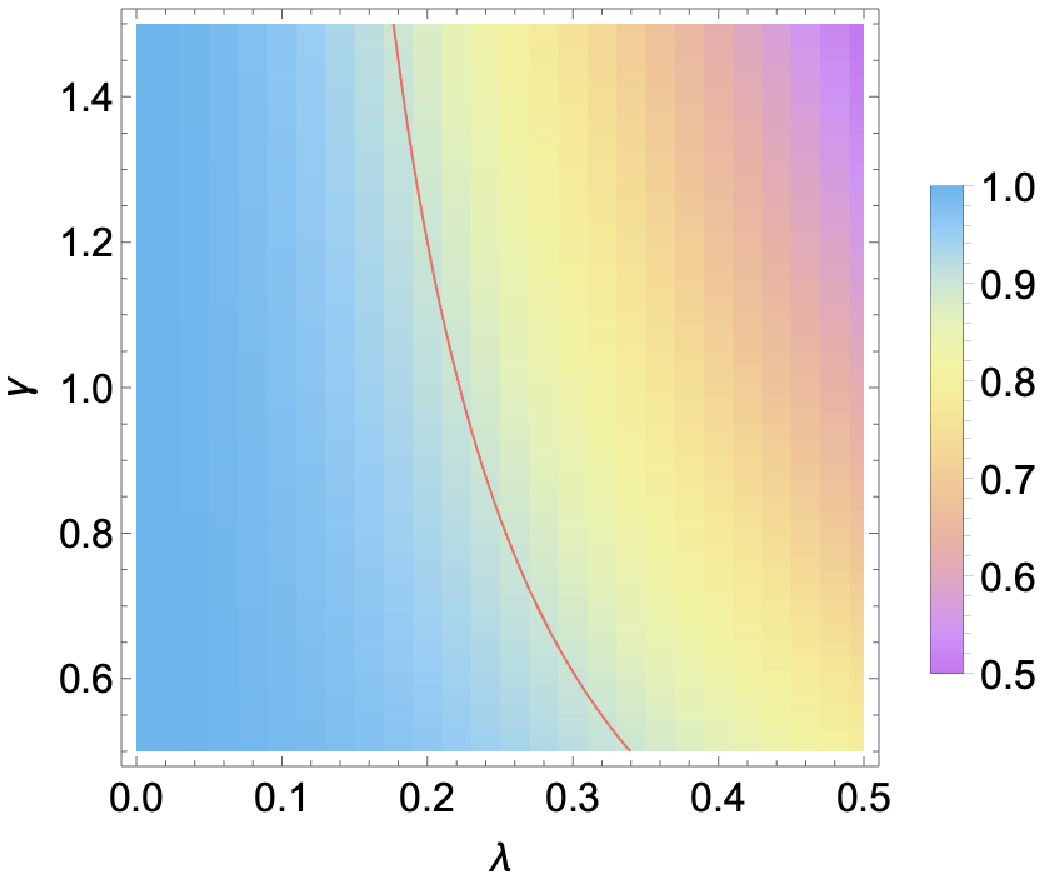, width=8cm}} 
  \fcaption{Fidelity between the target state (rotation by $45$ degrees at the final site) and the final evolved state under open system dynamics, as a function of the system-bath coupling $\lambda$ and bath memory parameter $\gamma$. The red line shows the boundary for fidelity $F=0.9$.}\label{fig_r45_ogl}
\end{figure}

In order to quantitatively study how the open system parameters affect the robustness of Krotov's method, we plot the fidelity at the final time between the evolved state and the target state, as a function of the system-bath coupling strength $\lambda$ and the bath memory parameter $\gamma$ in Fig.~\ref{fig_r45_ogl}. As is expected, the fidelity would decay faster with larger system-bath coupling strength. It is also interesting to note that in the region where the non-Markovian memory effect is stronger (smaller $\gamma$), the fidelity would drop slower when we increase the coupling strengths. It is demonstrated that the control is typically more robust in a non-Markovian regime. Note that for the particular memory function under consideration, smaller memory parameter $\gamma$ also lowers the effective system-bath coupling.

\section{Conclusion}\label{sec_sum}

Quantum control is a critical issue that real-world quantum technologies must address. In this paper, we have considered the optimal control of a quantum cavity array involving transfers of a Schr\"{o}dinger cat state through a linear harmonic oscillator chain and arbitrary rotations. We have shown that the quantum mechanical system can realize an effective autonomous control of its states by making internal time-dependent changes. Here, the control functions are represented by the time-dependent couplings and the oscillators' frequencies. We have demonstrated that Krotov's method may attain the optimal quantum control shapes for the system parameters. For real physical systems, the control functions cannot take arbitrary values - that is, they can neither be too large nor change too fast. For many physically interesting cases, the control functions can typically be realized within physically allowable values, because the control functions obtained by Krotov's algorithm are sensibly dependent on the initial guess, which can be chosen through trials such that the resulting control functions are well behaved. We note that for the model under consideration, in addition to the rotated cat states studied here, well-behaved control functions also exist for many other interesting states such as  entangled states $|\alpha\rangle|\alpha \rangle + |-\alpha\rangle|-\alpha \rangle$ and $N00N$ states $|N0 \rangle + |0N \rangle$ (up to a normalized factor).

To show the robustness of the quantum control of the quantum cat system, we have considered a more realistic situation where the quantum cat system is coupled to a generic noisy environment. In these more practical cases, we have shown that it is possible to achieve the final controlled target with high fidelity within the predetermined time. We have shown that the quantum control strategy adopted in this paper can be robust against environmental noises. It should be noted that here, we derived the controls in an ideal closed system setting and studied its robustness against open system effects. While it is possible to extend Krotov's method to vectorized Markov master equations~\cite{Goerz2019w}, deriving a consistent control in a full non-Markovian context would be a more challenging task since the dissipative terms would have an intricate dependence on the time-dependent control Hamiltonians. It would be interesting to implement our control method in a full non-Markovian setup or study how to use it consistently with other decoherence eliminating controls, such as  dynamical decoupling pulses~\cite{openc_bb} or leakage elimination operators~\cite{openc_leo}. It might also be of interest to compare the performances of our control method with alternative optimal control algorithms, such as~\cite{alt_optalgo1, alt_optalgo2, alt_optalgoML, alt_optalgoAD}.

We believe that the control method studied here is of interest for many quantum information processing tasks where continuous quantum state transfer and control are needed~\cite{Ralph2003y}. For example, it has been demonstrated that cat states can be used to deterministically encode quantum information~\cite{qcat_app1}, and offer a novel paradigm for universal quantum computation~\cite{qcat_app2}. An optical implementation for universal quantum logic gates and quantum metrology applications using superpositions of cat states has also been proposed~\cite{qcat_app3}. Recently, coupled cavity arrays have been realized on various platforms~\cite{cavarr_1}, such as photonic-bandgap metamaterial~\cite{cavarr_2}. The real time control of the harmonic oscillators may be realized with ion traps~\cite{ho_ion} or with circuit QED platforms~\cite{ho_rlc1, ho_rlc2}. It would also be of interest to test this protocol on quantum computers, such as superconducting quantum computers~\cite{qcomptr_sc} or even with the IBM Q using a continuous variable approach~\cite{qcomptr_ibmq}. The model may also be of interest for implementations of quantum entanglement generation in an array of optical cavities or a continuous-variable quantum network.

\nonumsection{Acknowledgements}

We thank Drs. Ting Yu and Rupak Chatterjee for the useful discussions. This project is partly supported by ART020-Quantum Technologies Project.  





\end{document}